\documentclass[preprint]{ptephy_v1}

\usepackage{graphicx, hyperref}
\usepackage{braket}
\usepackage{amsfonts,amssymb,bm,amsmath}

\newcommand{\paren}[1]{\left(#1\right)}
\newcommand{\Paren}[1]{\left[#1\right]}

\def\Tr{\mathrm{Tr}}
\def\tTr{\mathrm{tTr}}
\def\Re{\mathrm{Re}}
\def\intdU{\int \prod_{n,\mu} dU_{n, \mu}}
\def\Svtilde{\tilde{S_v}(U(n,\mu))}

\numberwithin{equation}{section}

\begin{document}

\title{Toward tensor renormalization group study of three-dimensional non-Abelian gauge theory}



\author[1]{Takaaki Kuwahara}
\affil{Graduate School of Science and Technology, Shizuoka University, 836 Ohya, Suruga-ku, Shizuoka 422-8529, Japan }

\author[1,2]{Asato Tsuchiya}
\affil{
  Department of Physics, Shizuoka University, 836 Ohya, Suruga-ku, Shizuoka 422-8529, Japan
  \email{kuwahara.takaaki.15@shizuoka.ac.jp, tsuchiya.asato@shizuoka.ac.jp}
}

\begin{abstract}
We propose a method to represent the path integral over gauge fields
as a tensor network.
We introduce a trial action with variational parameters and
generate gauge field configurations with the weight defined by the trial action.
We construct initial tensors with indices labelling these
gauge field configurations.
We perform the tensor renormalization group with the initial tensors
and optimize the variational parameters.
As a first step to the TRG study of
non-Abelian gauge theory in more than two dimensions,
we apply this method
to three-dimensional  pure SU(2) gauge theory.
Our result for the free energy agrees with the analytical results
in weak and strong coupling regimes.
\end{abstract}
\maketitle

\setcounter{footnote}{0}

\section{Introduction}
\setcounter{equation}{0}
\renewcommand{\thefootnote}{\arabic{footnote}}
Much attention has been paid to the tensor renormalization group (TRG)
\cite{Levin:2006jai}
as a new numerical method for studying
lattice field theories \cite{Liu:2013nsa,Xie_2012,Shimizu:2014uva,Shimizu:2014fsa,Shimizu:2017onf,Butt:2019uul,Takeda:2014vwa,Akiyama:2020soe,Kadoh:2018tis,Kadoh:2019ube,Akiyama:2020ntf,Akiyama:2021zhf,Asaduzzaman:2019mtx,Bazavov:2019qih,Fukuma:2021cni,Hirasawa:2021qvh, Kuramashi:2018mmi, Akiyama:2022eip,Kadoh:2018hqq,Kawauchi:2016xng,Akiyama:2020sfo,Adachi:2019paf,Kadoh:2019kqk,Kadoh:2021fri,Kuramashi:2019cgs}, since the
method is free from the sign problem and
enables us to take the large volume limit quite easily.

In the TRG, it is nontrivial to represent the path integral over continuous bosonic
fields as a tensor network that provides initial tensors, while
it is rather straightforward to represent that over fermionic fields as a tensor network.
For scalar fields, the Gauss-Hermite quadrature works well
in two \cite{Kadoh:2018tis,Kadoh:2019ube} and
four \cite{Akiyama:2020ntf,Akiyama:2021zhf} dimensions.
For gauge theories, the character expansion is successfully applied to
the U(1) gauge field
\cite{Shimizu:2014uva,Shimizu:2014fsa,Shimizu:2017onf,Kawauchi:2016xng,Kuramashi:2019cgs}, SU(2) gauge field \cite{Asaduzzaman:2019mtx,Bazavov:2019qih}, and
SU($N$) and U($N$) gauge fields \cite{Hirasawa:2021qvh} in two dimensions, while
a random sampling method is applied to the SU(2) and SU(3) gauge fields
in two dimensions \cite{Fukuma:2021cni}.
In the character expansion, the tensor indices correspond to the labels
that specify irreducible representations belonging to a subset
of all irreducible representations of a gauge group.
In the random sampling method, the tensor indices label
gauge configurations that are generated numerically with the Haar measure.

Moreover, the cost of calculation for the method is more sensitive to
the dimensionality of space-time than other methods such as the Monte Carlo method.
Indeed, in gauge theories in more than two dimensions,  it looks hard
to make the above subset in the character expansion large.
As for the random sampling method, we find that
it works well in the strong coupling regime for
three-dimensional pure SU(2) gauge theory.
However, we will see that it is not applicable to other regimes
when the range of the tensor indices is hard to increase.
Thus, as far as we know,
no non-Abelian gauge theories in more than two dimensions
have been studied through the TRG so far.
Hence, it is desirable to develop a more efficient method to represent
the path integral over gauge fields as a tensor network.

In this paper we propose a candidate for such a method.
We introduce a trial action with variational parameters
for a link variable and numerically generate gauge field configurations
with the weight defined by the trial action.
We construct initial tensors with indices labelling these
gauge field configurations.
We perform the tensor renormalization group with the initial tensors for
various values of the variational parameters, and fix the variational parameters
such that
the result is insensitive to them in the spirit of the mean field approximation and
the Gaussian expansion method (improved mean field approximation or delta expansion;
see, for instance, Ref.\cite{Nishimura:2003gz} and references therein).
Our method can be viewed as an improvement of the random sampling method
\cite{Fukuma:2021cni}.
As a first step to the TRG study of non-Abelian gauge theory in more than two dimensions,
we apply this method to three-dimensional pure SU(2) gauge theory.
We find that the result for the free energy agrees with the analytical results
in the weak and strong coupling regimes.

This paper is organized as follows.
In Sect. 2 we describe our method to represent the path integral over gauge fields
as a tensor network.
In Sect. 3, we show the result for three-dimensional pure SU(2) gauge theory obtained
using our method.
Sect. 4 is devoted to conclusion and discussion.
In the appendix, the construction of initial tensors is explained in detail.

\section{Tensor network formulation}
\label{sec: TN formulation}
\setcounter{equation}{0}
In this section we explain our method to
represent three-dimensional pure SU($N$) gauge theory on the lattice
as a tensor network. To extend this to higher dimensions is straightforward.

The partition function is defined by
\begin{equation}
  Z = \intdU e^{-S} \ ,
\end{equation}
where $n$ are the lattice sites, and $(n,\mu)$ with $\mu=1,2,3$ specify the links.
$U_{n,\mu}$ are the link variables that take SU($N$) matrices,
and $dU_{n,\mu}$ are the Haar measure
normalized as $\int dU_{n,\mu} = 1$.
The plaquette action $S$ is defined by
\begin{equation}
  S =  \frac{\beta}{N} \sum_{n, \mu > \nu} \Re \Tr (1-U_{\mu\nu}(n))
\end{equation}
with $U_{\mu\nu}(n) = U_{n,\mu} U_{n +\hat{\mu}, \nu}
U_{n+\hat{\nu}, \mu}^\dagger U_{n, \nu}^\dagger$.

Here we introduce a trial action $S_v$ with some variational parameters such that
the partition function is unchanged:
\begin{equation}
Z = \intdU e^{-(S - S_v) - S_v} \ .
\end{equation}
We assume that $S_v$ is given by the sum over single link actions as
\begin{equation}
  \begin{split}
    S_v = \sum_{n, \mu} \tilde{S}_v(U_{n,\mu}) \\
  \end{split}
\end{equation}
and that the partition function for the single link action $\tilde{S}_v$,
\begin{align}
Z_v = \int dU e^{-\Svtilde}   \ , \nonumber
\end{align}
is calculable by a certain method.
%
The simplest example of $\tilde{S}_v$ is given by
\begin{align}
\tilde{S}_v(U)= -\frac{H}{N}\Re\Tr U \ ,
\label{example of trial action}
\end{align}
where $H$ is a variational parameter.
In the SU(2) case, $Z_v$ is calculated as
\begin{align}
Z_v=2\frac{I_1(H)}{H}\ ,
\end{align}
where $I_1$ is the modified Bessel function.
Later, we will use (\ref{example of trial action}) for SU(2).

Then, we represent $Z$ as
\begin{align}
Z = Z_v^{3V} \langle e^{-(S-S_v)} \rangle_v \ ,
\end{align}
where $V$ is the number of sites, and $\langle \cdots \rangle_v$ stands for
the statistical average with respect to the Boltzmann weight $e^{-S_v}$:
\begin{align}
\langle \cdots \rangle_v = \frac{1}{Z_v^{3V}}\int \prod_{n,\mu} dU_{n,\mu} \cdots e^{-\sum{n,\mu}\tilde{S_v}(U_{n,\mu})} \ .
\end{align}

We generate $K$ configurations of $U$ with the Boltzmann weight $e^{-\tilde{S}_v(U)}$
in general numerically and approximate the integral over each of $U_{n,\mu}$ as
\begin{equation}
    \int dU_{n,\mu} g(U_{n,\mu},U_{n',\mu'},\ldots) \approx \frac{1}{K} \sum_{i=1}^K g(U_i,U_{n',\mu'},\ldots) \ ,
\end{equation}
where $U_i$ are elements of the set $ G = \{U_1, U_2, \cdots, U_K\}$.
We use the labels of the configurations $i$ as the tensor indices.

In principle, the calculation is independent of $\tilde{S}_v$
if  one can make $K$  large enough in performing
the TRG. For instance, one can take $\tilde{S}_v=0$, which corresponds to
producing configurations randomly with the Haar measure.
This corresponds to the random sampling method that is
used for a tensor network representation of two-dimensional pure gauge theory \cite{Fukuma:2021cni}\footnote{
It is shown in this case that $K$ is allowed to be small.
}.
However, it is difficult to make $K$ large
due to the cost of calculation. Indeed, it turns out in Sect. 3 that in
pure SU(2) gauge theory
$\tilde{S}_v=0$ with reasonable values of $K$ works only in the strong coupling regime.
In practice, we need to sample
configurations such that the TRG works efficiently. We choose
$\tilde{S}_v$ appropriately and optimize the variational parameters such that
the result is insensitive to them.

\begin{figure}[t]
  \begin{minipage}[b]{0.55\linewidth}
    \centering
    \includegraphics[keepaspectratio, scale=0.4]{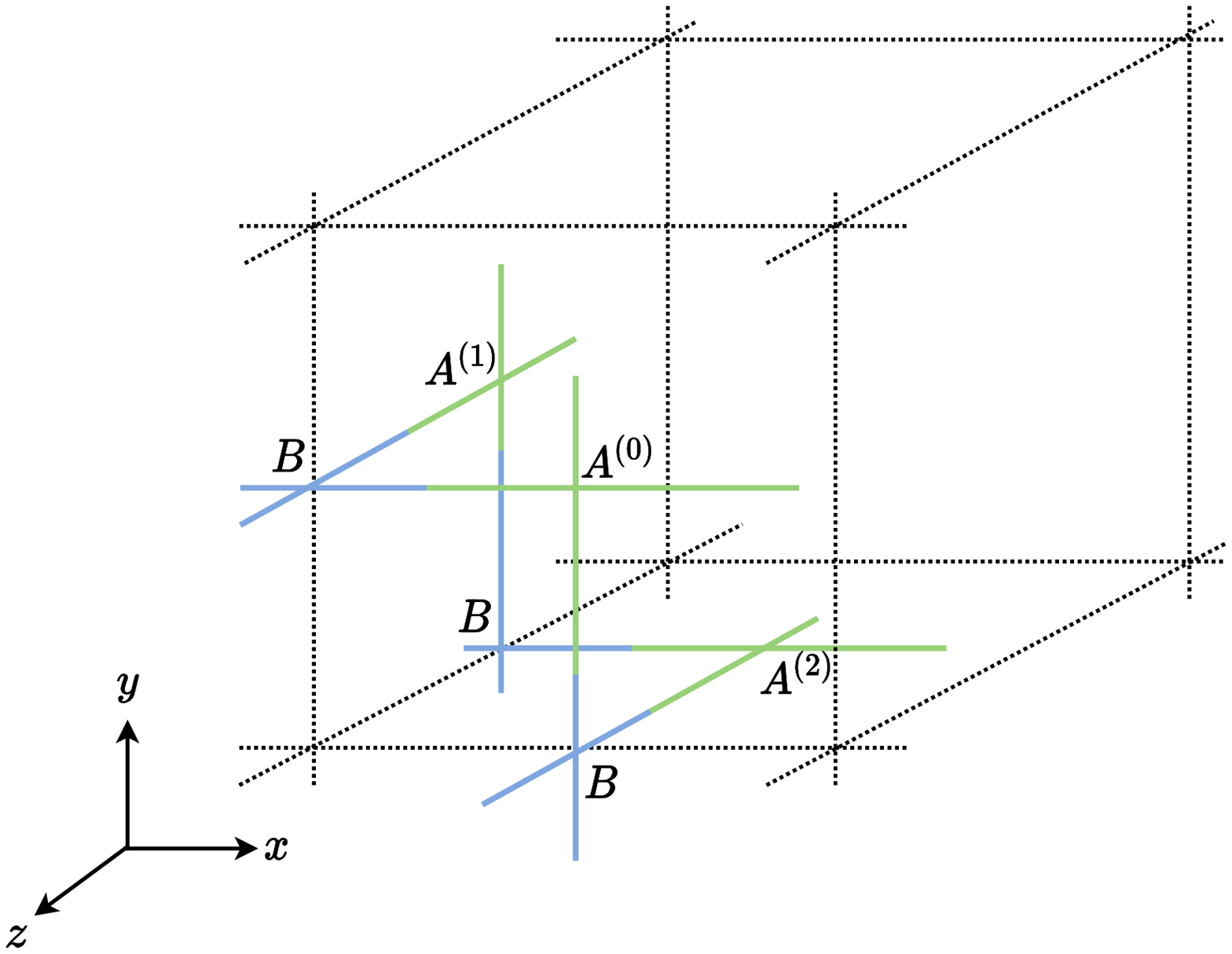}
  \end{minipage}
  \begin{minipage}[b]{0.4\linewidth}
    \centering
    \includegraphics[keepaspectratio, scale=0.5]{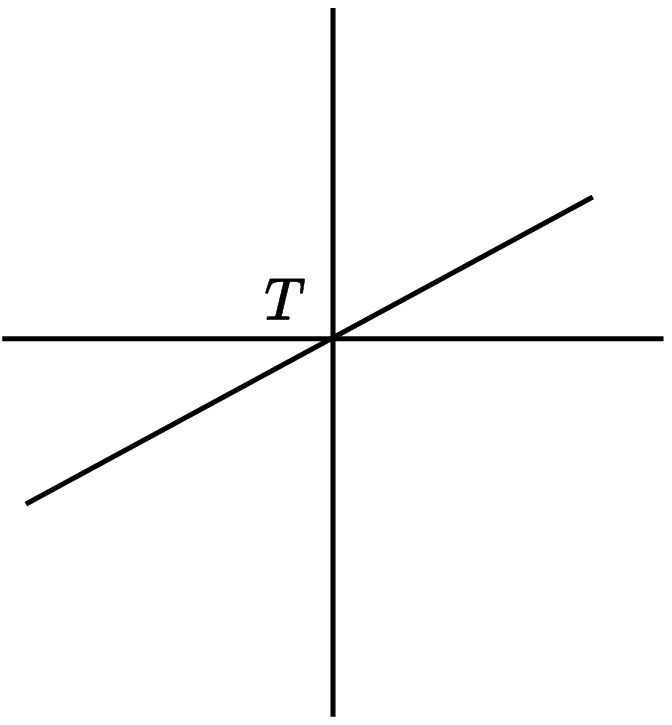}
  \end{minipage}
  \caption{
    (Left) $A$ tensors on plaquettes and $B$ tensors on links.
    (Right) $T$ tensor in the center of a cube,
    where $T =A^{ (0) } \otimes A^{ (1) } \otimes A^{ (2) } \otimes B \otimes B\otimes B$.
  }
  \label{fig: initial tensors}
\end{figure}

We construct a tensor that resides on the center of a plaquette:
\begin{equation}
  \label{A tensor}
  A_{ijkl} = \exp \Paren{\frac{\beta}{N}  \Tr \paren{U_i U_j U_k^\dagger U_l^\dagger} - \frac{1}{4} \paren{ \tilde{S}_v(U_i) + \tilde{S}_v(U_j) + \tilde{S}_v(U_k) + \tilde{S}_v(U_l)}} \ .
\end{equation}
We introduce a tensor $B_{ijkl}$ to construct a six-rank tensor
from $A_{ijkl}$, following the exact blocking formula\cite{Liu:2013nsa}.
$B_{ijkl}$ are placed on links, and take the form
\begin{equation}
  B_{ijkl} = \delta_{ijkl} = \delta_{ij}\delta_{jk}\delta_{kl}\delta_{li} \ .
\end{equation}
A graphical representation of $A$ tensors and $B$ tensors is given in Fig.~\ref{fig: initial tensors}. By using $A$ and $B$ tensors, the initial tensor $T$ is constructed as
\begin{equation}
  \label{exact initial T form}
  T =A^{ (0) } \otimes A^{ (1) } \otimes A^{ (2) } \otimes B \otimes B\otimes B \ .
\end{equation}
Here we generate three configuration sets $G^{(0)}$, $G^{(1)}$, and $G^{(2)}$ for $A^{(0)}$, $A^{(1)}$, and $A^{(2)}$, respectively,
to improve the K dependence\cite{Fukuma:2021cni}
\footnote{
  We can further consider  a couple of three configuration sets,
$\{ G^{(0)}, G^{(1)}, G^{(2)} \}$ and $\{ G'^{(0)}, G'^{(1)}, G'^{(2)} \}$, each of which is
used on even/odd sites~\cite{Fukuma:2021cni}.
}.
The $A^{(0)}$, $A^{(1)}$, and $A^{(2)}$ tensors are defined on the ($xy$), ($yz$), and ($zx$) planes, respectively(see Fig.~\ref{fig: initial tensors}), while the
$T$ tensor is a six-rank tensor which is  placed in the center of a cube and
whose bond dimension is $K^2$.
Thus,  we obtain a tensor network representation of the partition function
\begin{equation}
  Z(K) = \paren{ e^{-\beta}\frac{Z_v}{K} }^{3V}  \tTr \otimes_{n} T  \ ,
\end{equation}
where $\tTr$ stands for the trace over tensor.

In what follows, we consider the SU(2) case and adopt Eq. (\ref{example of trial action})
as the trial action.
We truncate the bond dimension for $T$ to $D$ by introducing isometries
as in the higher-order TRG(HOTRG) method~\cite{Xie_2012}.
The truncation procedure is summarized in Appendix \ref{appendix: construction of the initial tensor}.

\section{Numerical results}
\setcounter{equation}{0}
In this section we show the numerical results for three-dimensional
SU(2) pure gauge theory on the lattice.
We calculate the free energy (density) $F = (1/V) \log Z$
by using our formulation introduced in the previous section
and the anisotropic TRG~\cite{Adachi:2019paf}.
We adopt Eq. (\ref{example of trial action}) with $N=2$  as the trial action $\tilde{S}_v$.
In the following results, the lattice size $L$, which is related to $V$ as $V=L^3$,
is fixed to $L = 1024$.

First, by using the Monte Carlo method with the Boltzmann weight $e^{-\tilde{S}_v}$,
we generate three sets of $K$ field configurations
$G^{(i)} = \{ U_1^{(i)}, U_2^{(i)}, \cdots, U_K^{(i)}\} \;  (i=0,1,2)$.
Second, we construct the $A^{(i)}$ tensors ($i=0,1,2$) from $G^{(i)}$ as
explained in the
previous section.
Third, by installing isometries to truncate the bond dimension from $K^2$ to $D$
as explained in Appendix \ref{appendix: construction of the initial tensor},
we construct the initial tensor $T$.
Finally, we apply the ATRG~\cite{Adachi:2019paf} with the bond dimension $D$
to the initial tensor $T$
to calculate the free energy $F=\frac{1}{V}\log Z$.
We perform the calculation of the free energy for various values of $H$ for fixed $\beta$
and search for a plateau of the free energy
under the change of $H$ because the free energy is originally
independent of $H$.

The estimates of the free energy have statistical errors in addition to
the systematic errors coming from the finiteness of $K$ and the bond dimension $D$.
The statistical errors that are given below as error bars are obtained from ten independent trials.

\subsection{The $D$ and $K$ dependencies}
In this subsection we examine the dependence of the free energy on $D$ and
$K$.
We choose $\beta=1$ and $\beta=50$ as typical values of small and large $\beta$,
respectively.

\begin{figure}[t]
  \begin{minipage}[t]{0.5\columnwidth}
    \centering
    \includegraphics[keepaspectratio, width=0.95\columnwidth]{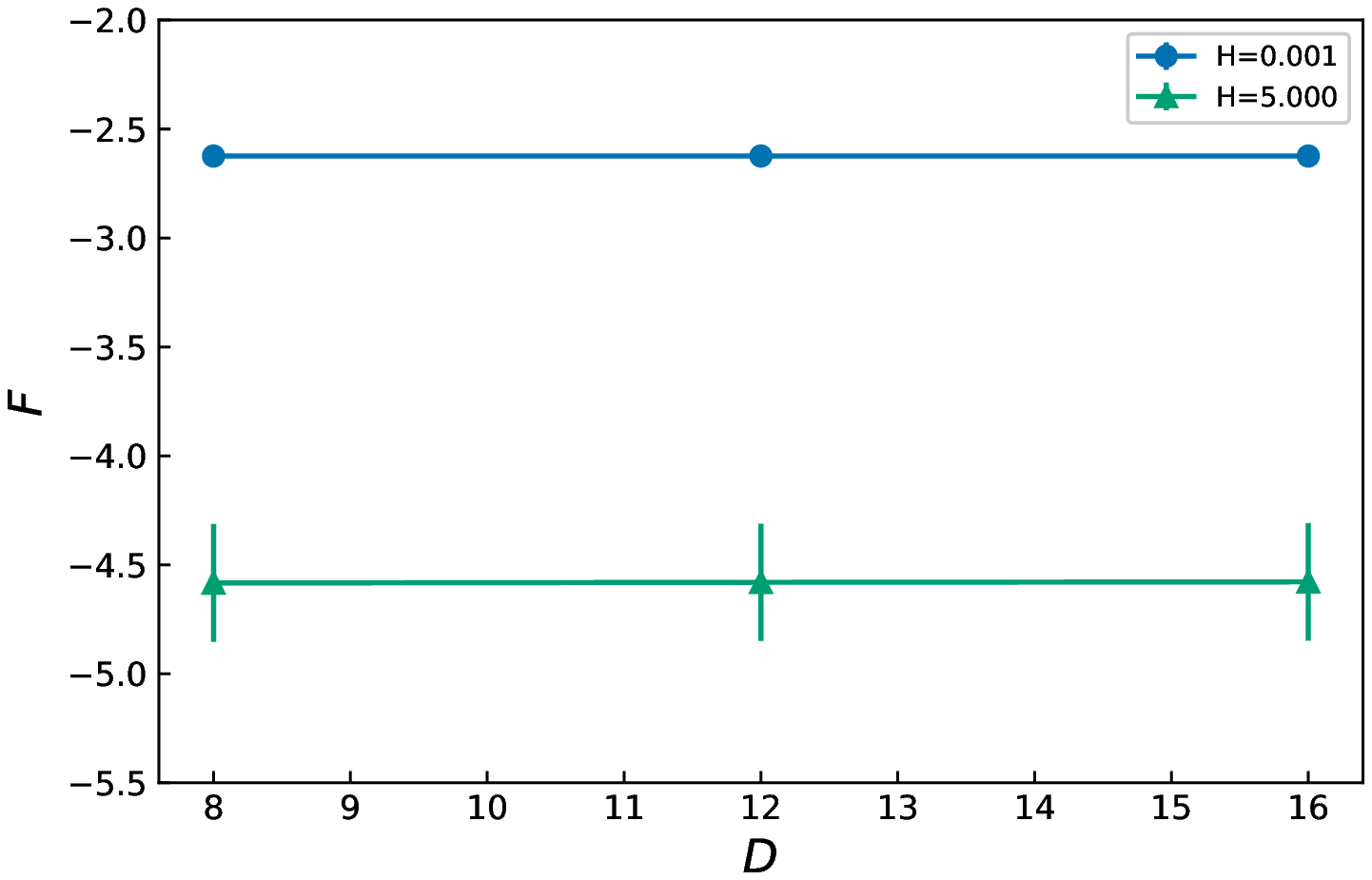}
  \end{minipage}
  \begin{minipage}[t]{0.5\columnwidth}
    \centering
    \includegraphics[keepaspectratio, width=0.95\columnwidth]{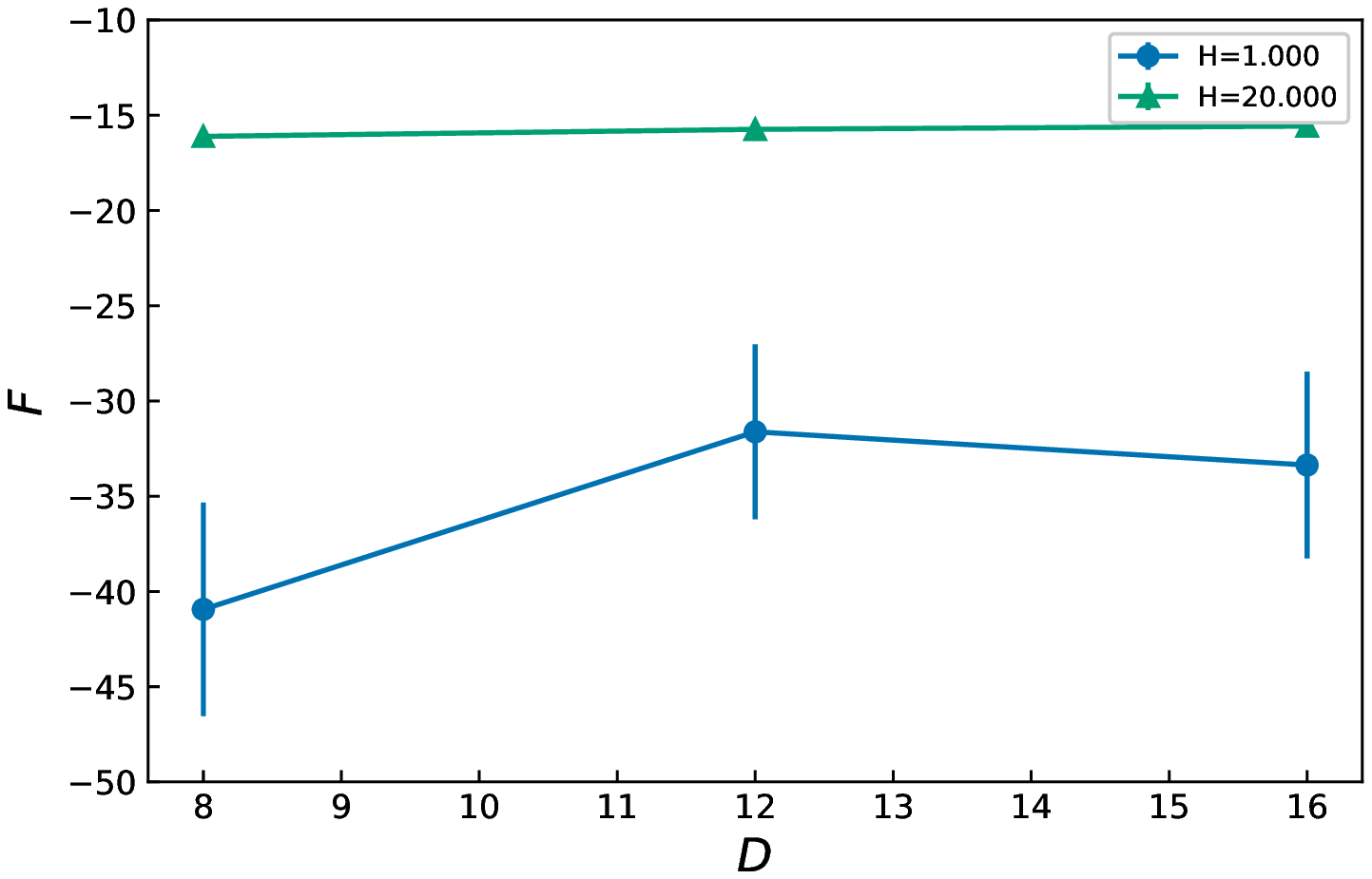}
  \end{minipage}
  \caption{
    The $D$ dependence of the free energy with
    $\beta=1$ (left) and $\beta=50$ (right), where $K=12$.
    The lines are drawn to guide the eye.
    (Left) The dots and triangles represent the results for $H=0.001$ and $H=5$,
              respectively. The statistical errors for $H=0.001$ are smaller than the symbol
             size.
    (Right) The dots and triangles represent the results for $H=1$ and $H=20$,
              respectively. The statistical errors for $H=20$ are smaller than the symbol size.
  }
  \label{fig: D dep}
\end{figure}

First, we examine the $D$ dependence.
The $D$ dependence of the free energy with $\beta=1$ and $\beta=50$
is shown in the left and right panels of Fig.~\ref{fig: D dep}, respectively.
Here $K$ is fixed to $K=12$, and the results for
two typical values of $H$ are shown.
We see in Fig.~\ref{fig: D dep} (left) that
the statistical errors for $H=0.001$ are much smaller than those for $H=5$,
and that the results for both $H=0.001$ and $H=5$ are stable against the change of $D$.
We see in Fig.~\ref{fig: D dep} (right) that
the statistical errors for $H=20$ are much smaller than those for $H=1$, and the result for $H=20$ is stable against the change of $D$ while that for $H=1$ is not.
These results imply that it is crucial in our algorithm to tune $H$
appropriately. In particular,
$D=12$ is considered to be sufficient in both the weak and strong
coupling regimes if $H$ is chosen appropriately.

\begin{figure}[th]
  \begin{minipage}[b]{0.5\columnwidth}
    \centering
    \includegraphics[keepaspectratio, width=0.95\columnwidth]{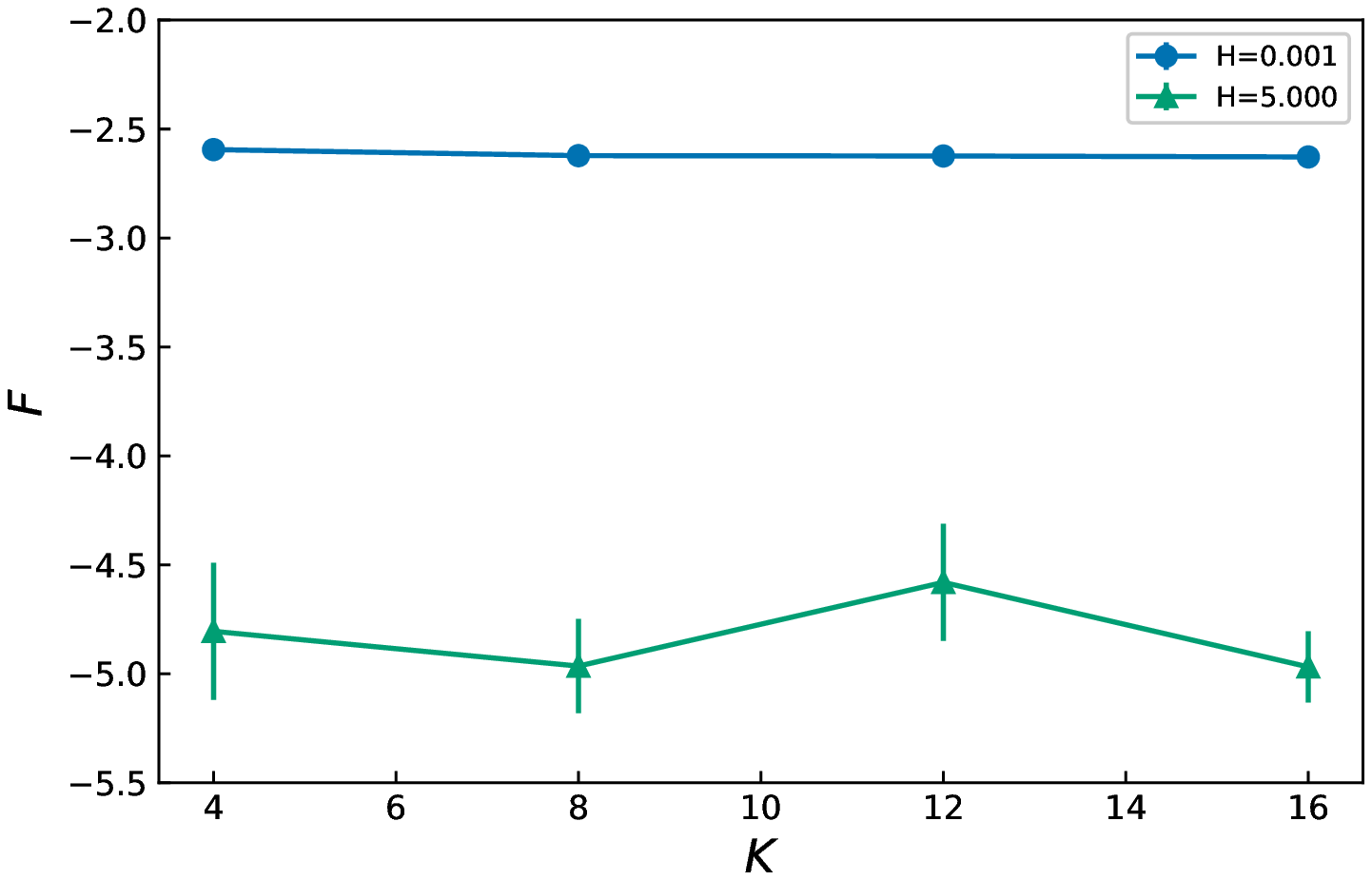}
  \end{minipage}
  \begin{minipage}[b]{0.5\columnwidth}
    \centering
    \includegraphics[keepaspectratio, width=0.95\columnwidth]{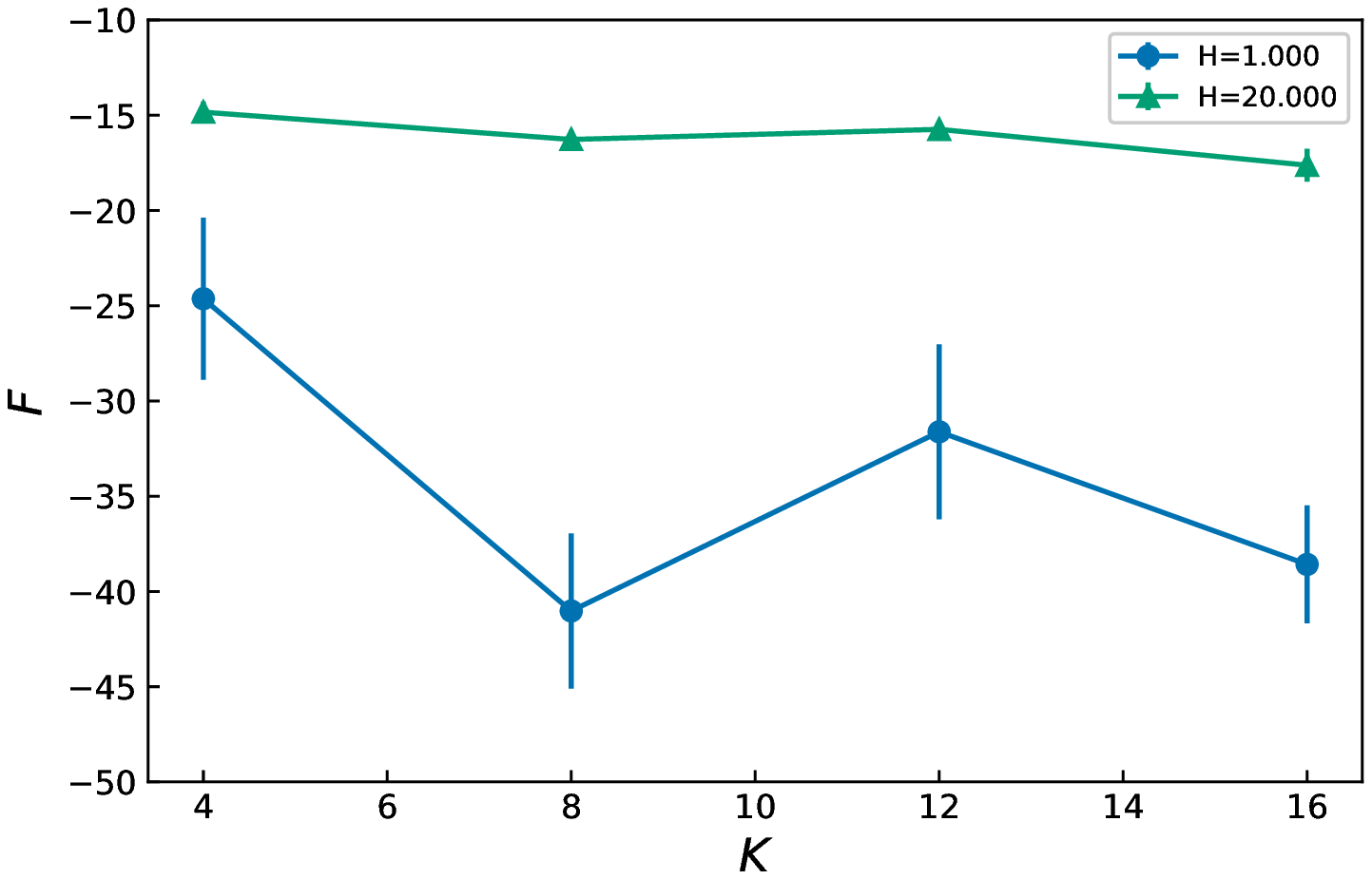}
  \end{minipage}
  \caption{The $K$ dependence of the free energy with $\beta=1$ (left) and
              $\beta=50$ (right), where $D=12$. The lines are drawn to guide the eye.
    (Left)  The dots and triangles represent the results for $H=0.001$ and $H=5$,
              respectively. The statistical errors for $H=0.001$ are smaller
               than the symbol size.
    (Right) The dots and triangles represent the results for $H=1$ and $H=20$,
              respectively. The statistical errors for $H=20$ are smaller than the symbol size.
  }
  \label{fig: K dep}
\end{figure}

Next, we examine the $K$ dependence.
The $K$ dependence of the free energy with $\beta=1$ and $\beta=50$
is shown in the left and right panels of Fig.~\ref{fig: K dep}, respectively.
Here, $D$ is fixed to $D=12$, and the results for
two typical values of $H$ are shown.
We see in Fig.~\ref{fig: K dep} (left) that
the statistical errors for $H=0.001$ are much smaller than those for $H=5$,
and that the result for $H=0.001$ is stable against the change of $K$ while
that for $H=5$ is not.
This again implies that tuning $H$ is crucial in our algorithm,
and $K=16$ is sufficient in the strong coupling regime.
Similarly, we see in Fig.~\ref{fig: K dep} (right) that
the statistical errors for $H=20$ are much smaller than those for $H=1$.
However, the result for $H=20$ does not look completely stable against the change of $K$
in the range $K \leq 16$.
Due to the limitation of available memory,
we take $K=16$ in the following calculations.
Indeed, as we will show in Sect.~\ref{sec: free energy},
the result for the free energy for $20\leq\beta\leq50$ agrees with the weak coupling expansion.
Thus, the $K$ dependence for $K\geq 16$ with $H \sim 20$ is expected not to be
large in the weak coupling regime.
From the above results, we set $D=12$ and $K=16$
in the following calculations.

\subsection{Free energy}
\label{sec: free energy}
\begin{figure}[thbp]
  \centering
  \includegraphics[width=0.8\linewidth]{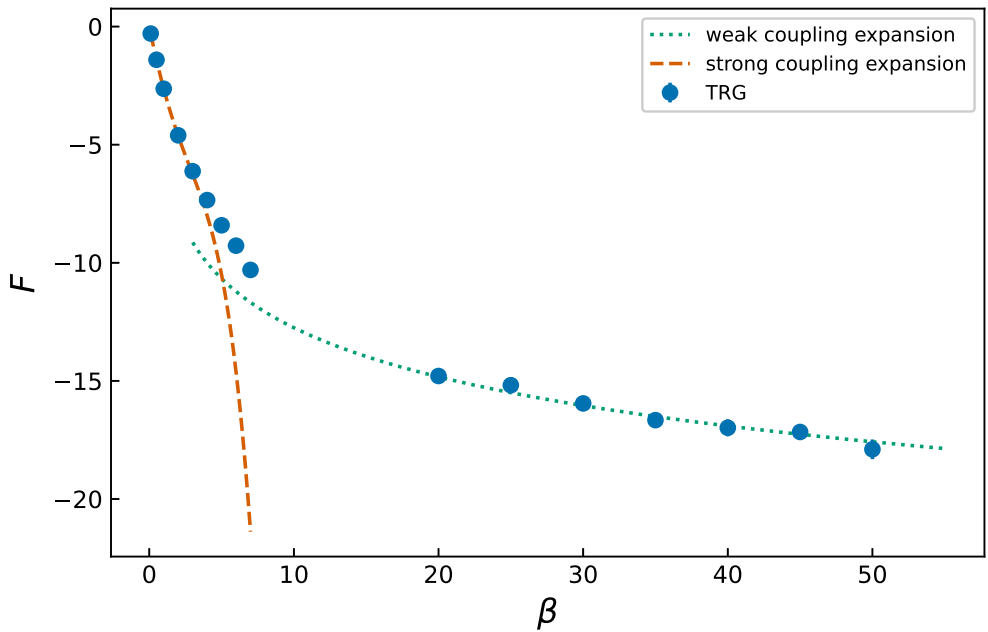}
  \caption{The free energy is plotted against $\beta$. The statistical errors are smaller
              than the symbol size. The strong coupling expansion is expressed by the dashed line,
              while the weak coupling expansion by the dotted line.
  }
  \label{fig: free energy}
\end{figure}

\begin{figure}[htbp]
  \begin{minipage}[b]{0.5\columnwidth}
    \centering
    \includegraphics[width=0.9\linewidth]{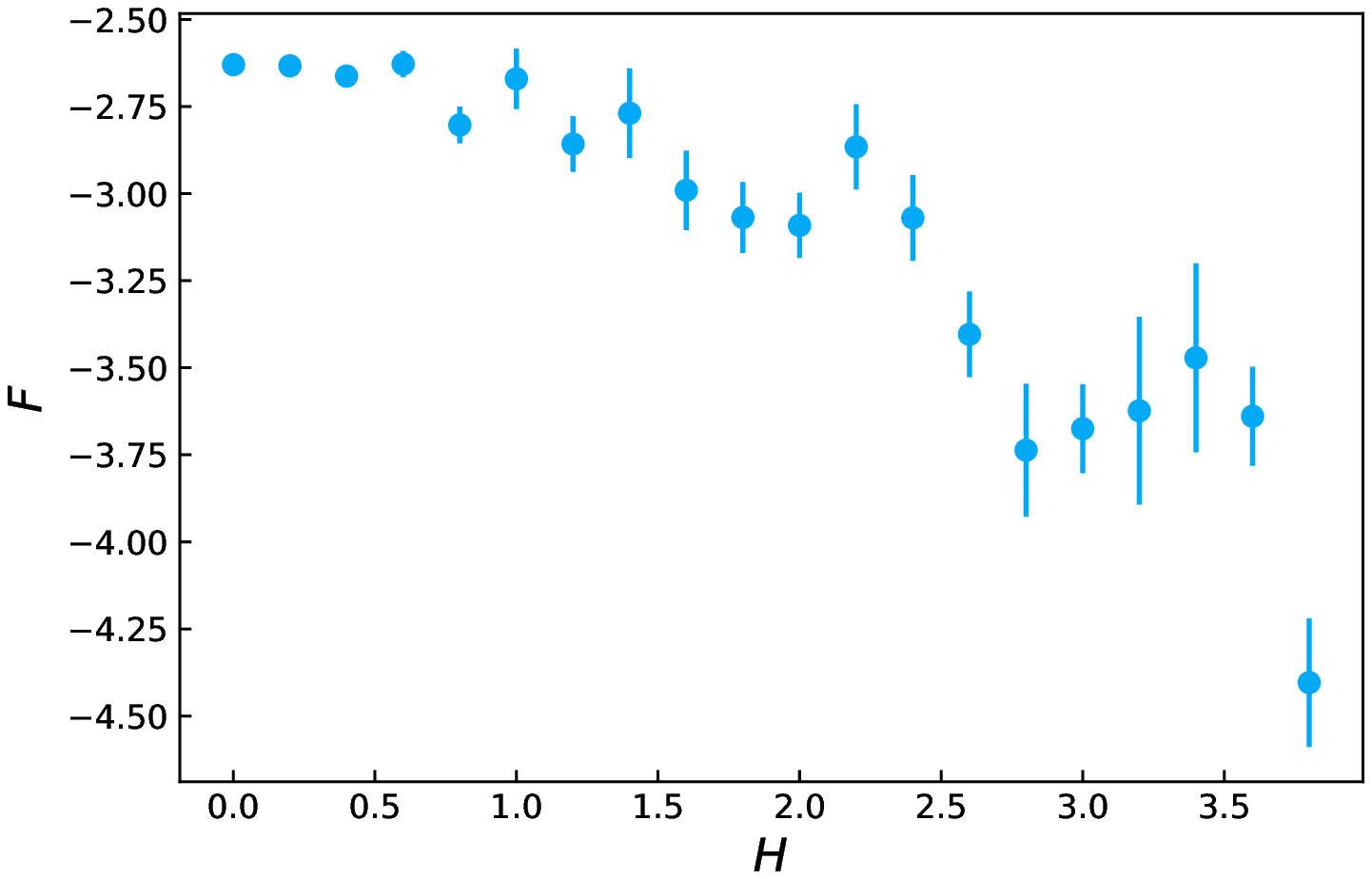}
  \end{minipage}
  \begin{minipage}[b]{0.5\columnwidth}
    \centering
    \includegraphics[width=0.9\linewidth]{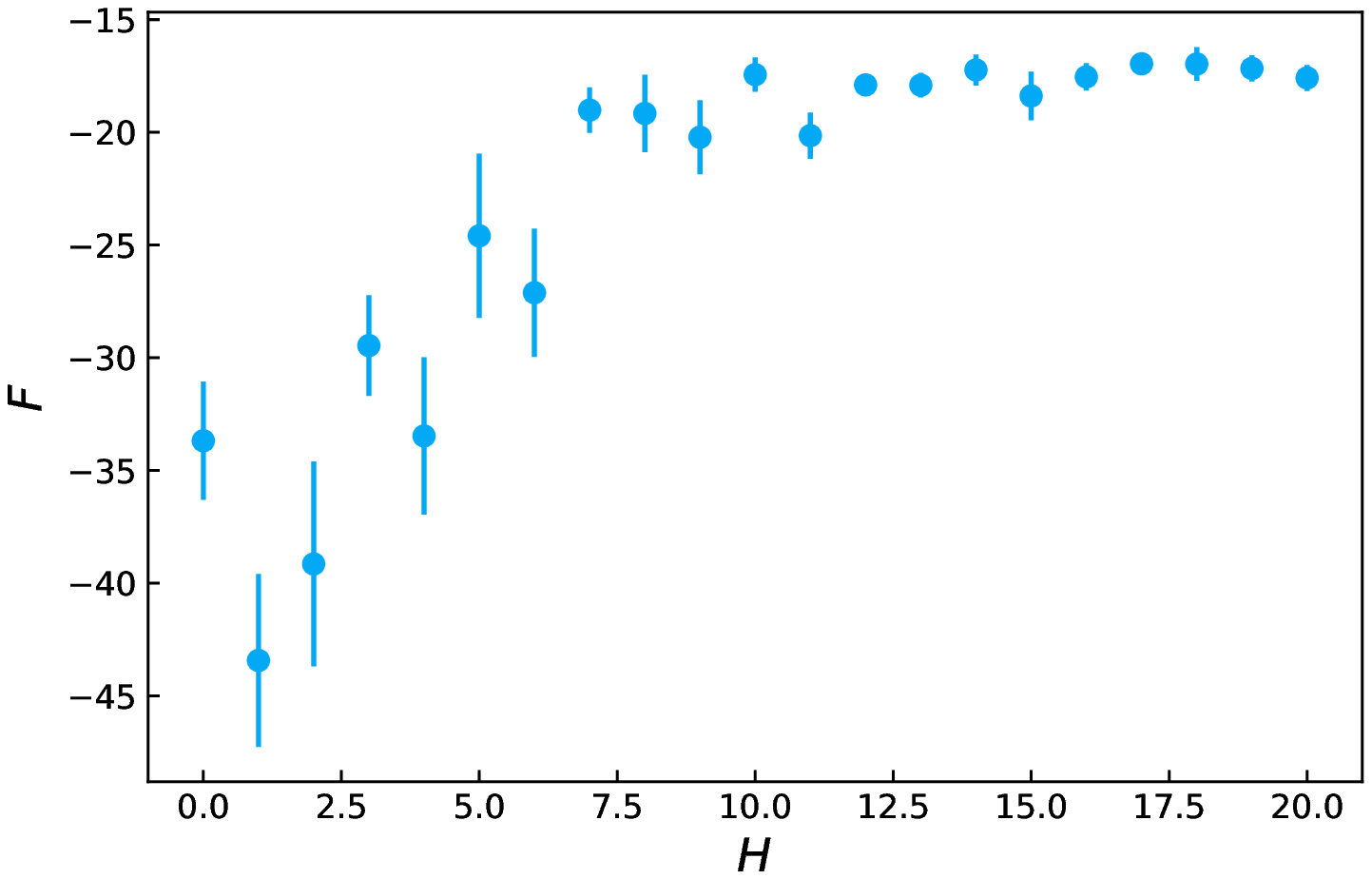}
  \end{minipage}
  \caption{The $H$ dependence of the free energy with $K=16$ and $D=12$
     for $\beta=1$(left) and $\beta=50$(right).
  }
  \label{fig: H dep}
\end{figure}

We show the result for the free energy in Fig.\ref{fig: free energy}.
Here $D$ and $K$ are fixed to
$D=12$ and $K=16$ as mentioned in the previous subsection.
We search for a plateau for each value of $\beta$
in the $0 < H \leq 20$ region.
The free energy is obtained from $F = F(H_*)$,
where $H_*$ has the smallest statistical error among the plateau.
Note that $H_*$ depends on $\beta$.
The dependence of the free energy on $H$  is shown in Fig.~\ref{fig: H dep}, where
we choose $\beta=1$ and $\beta=50$ as typical small and large values of $\beta$,
respectively.
We see that there is a plateau in the $H \leq0.6$ region for $\beta=1$ and
in the $H\leq16$ region for $\beta=50$.
We take $H_* = 0.001$ in $\beta \leq 7$ and $H_* > 10$ in $\beta \geq 20$.
($H=0$ should also work for $\beta \leq 7$.)

The strong coupling expansion of the free energy is given by
\begin{equation}
  F(\beta) = -3 \beta + \frac{3}{8}  \beta^2 - \frac{3}{384}  \beta^4 + \mathcal{O}(\beta^6) \ ,
\end{equation}
which is expressed by the dashed line
 in Fig.~\ref{fig: free energy}.
The weak coupling expansion of the expectation value of a plaquette is given by
$W_{1 \times 1} = e^{-1/\beta}$~\cite{Muller:1980hz}.
Thus we have
\begin{equation}
  F(\beta) =  - 3\log\beta  + C + \mathcal{O}\paren{ \frac{1}{\beta} }
\label{weak coupling expansion}
\end{equation}
with $C$ being an integration constant.
We determine the constant $C$ as $C = - 5.8426$ by fitting the data
in the $20 \leq \beta \leq 50$ region to $- 3\log\beta  + C$.
The weak coupling expansion is expressed by the dotted line in Fig.~\ref{fig: free energy}.
The result indeed agrees with the strong and weak coupling expansion,
in the strong and weak regimes, respectively.
However, in the $7 \leq \beta \leq 19$ region, we cannot find any definite plateau.
We expect this to be resolved by increasing $K$ and/or improving the trial action.

Our result suggests that the random sampling method \cite{Fukuma:2021cni} works
 in the strong coupling regime in higher-dimensional gauge theories.
If $K$ cannot be made large enough,
another method is needed in the intermediate and weak coupling regimes.
Our method is a candidate for such a method.

\section{Conclusion and discussion}
\setcounter{equation}{0}
We proposed a method to represent the path integral over
gauge fields as a tensor network.
In our method, tensor indices label gauge field configurations that are
generated with the weight determined by the trial action with variational parameters.
We construct initial tensors with these indices and perform
the TRG with the initial tensors for various values of the variational parameters
to fix the variational parameters such that the result is insensitive to them.
As a first step to the TRG study of non-Abelian gauge theories in more than two dimensions,
we studied three-dimensional  pure SU(2) gauge theory by using our method with
the ATRG.
We reproduced the weak and strong coupling behaviors of the free energy.
We found that the random sampling method (corresponding to $H=0$)
works in the strong coupling regime, while
tuning $H$ to a nonzero value is needed in the weak coupling regime.
Our result suggests that our method can be used for studying
gauge theories in more than two dimensions.

It is likely that we need to perform the calculation with larger $K$ and/or to
improve the trial action to see complete stability of the free energy against
the change of $K$ in the weak and intermediate coupling regime.
and find plateaus in the intermediate coupling regime \footnote{We should also try to
introduce a couple of three configuration sets, each of which is used on even/odd
sites.}

In order to establish the effectiveness of our method, we should study the physics of three-dimensional SU(2) gauge theory such as the string tension and the finite-temperature phase transition\cite{Kuramashi:2018mmi}. Furthermore, inclusion of matter, topological terms, the chemical potential, extension to other non-Abelian gauge groups, and extension to four dimensions are left as future work. We hope that our method will indeed be powerful for problems with complex actions.

\section*{Acknowledgments}
We would like to thank S. Akiyama, D. Kadoh and S. Takeda for discussions on the TRG.
The computation was carried out using the
supercomputer ``Flow'' at Information Technology Center, Nagoya University.
A.T. was supported in part by Grant-in-Aid for Scientific Research (Nos. 18K03614 and 21K03532) from
Japan Society for the Promotion of Science.

\appendix

\section{Construction of the initial tensor}
\label{appendix: construction of the initial tensor}

\begin{figure}[t]
  \centering
  \includegraphics[width=0.5\linewidth]{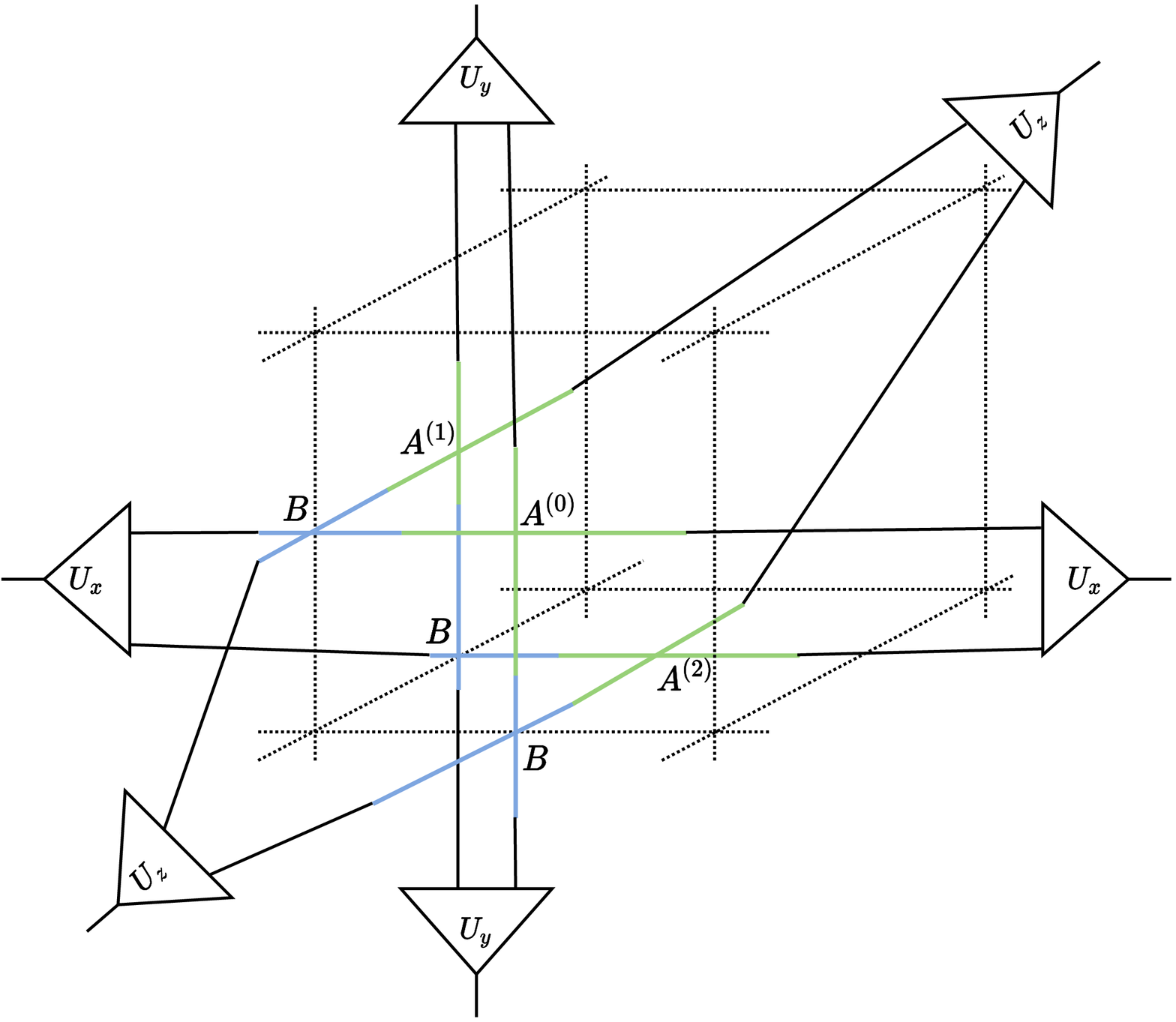}
  \caption{
    Isometries $U_x$, $U_y$, and $U_z$ for $x$, $y$, and $z$ directions, respectively.
  }
  \label{fig: HOSVD}
\end{figure}

\begin{figure}[t]
  \centering
  \includegraphics[width=0.5\linewidth]{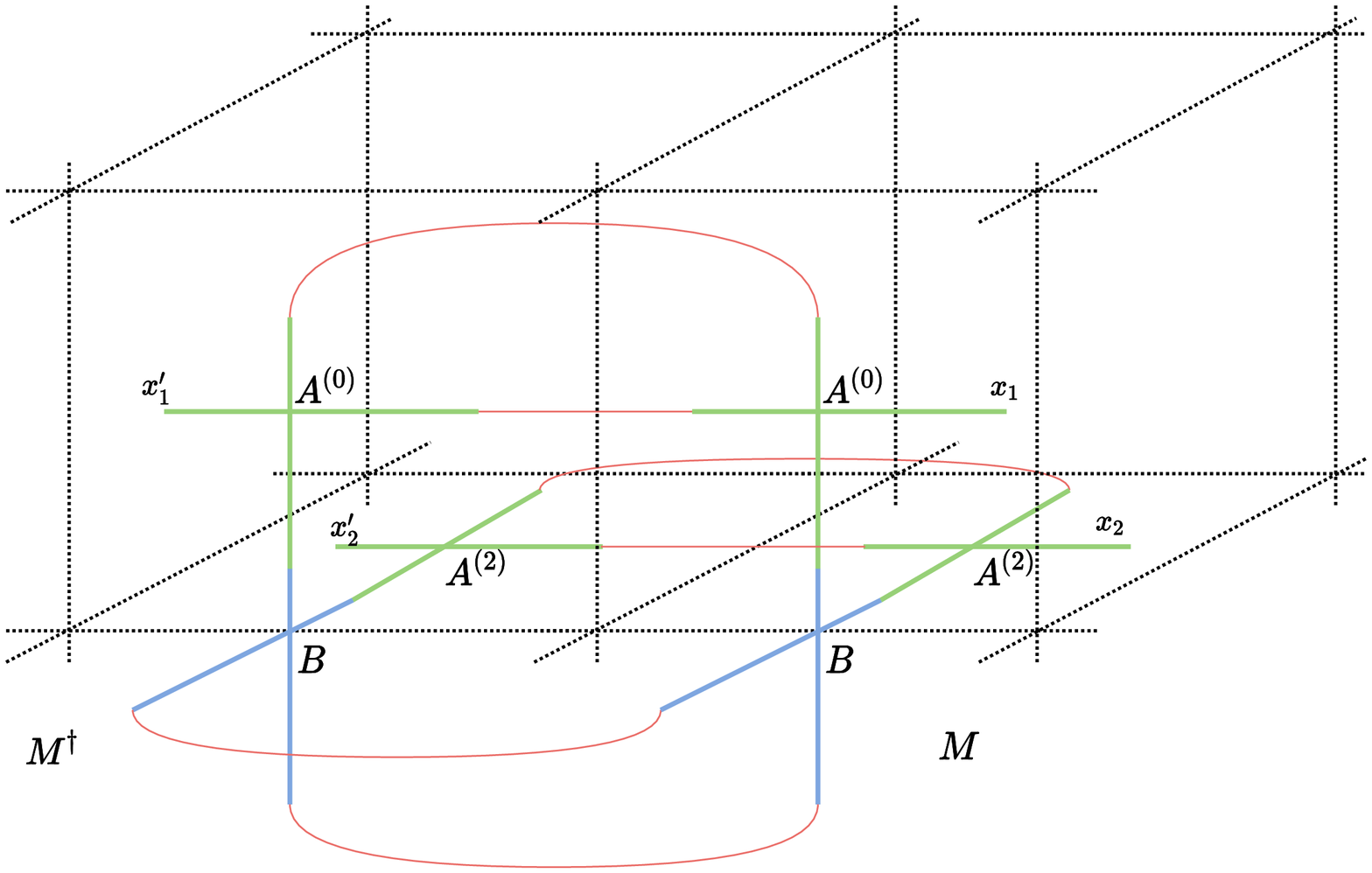}
  \caption{
    Coarse-graining for the $x$ direction.
    We consider the indices of $A^{(0)}$ and $A^{(2)}$, $(x_1, x_2)$, as the rows
    of the matrix $M$.
    We calculate $M M^\dag$ by making the contractions expressed by the red lines.
  }
  \label{fig: CG for x}
\end{figure}

In this appendix we describe the details of the construction of the initial tensor.
We have three $A$ tensors $\{A^{(0)},  A^{(1)}, A^{(2)}\}$ and three $B$ tensors that were introduced in Sect. \ref{sec: TN formulation}.
If the initial $T$ tensor is constructed exactly,
the six-rank tensor needs an $\mathcal{O}\paren{(K^2)^6}$ memory footprint.
For this reason, we install isometries to reduce the bond dimension from $K^2$ to $D$.
We apply HOTRG\cite{Xie_2012} to coarse-grain the $x$, $y$, and $z$ directions as shown in Fig.~\ref{fig: HOSVD}.
First, we introduce the isometries for the $x$ direction.
We perform higher-order singular value decomposition for $M \equiv A^{(0)} \otimes A^{(1)} \otimes B$.
$M$ is a matrix whose rows consist of the indices of $A^{(0)}$ and $A^{(2)}$ corresponding
to the right side (see Fig.~\ref{fig: CG for x}) and the columns consist of the other indices
 (see Fig.~\ref{fig: CG for x}) .
Then, we calculate $MM^{\dagger}$, which is a Hermitian matrix, and
perform the canonical transformation of $M M^\dagger$ as
\begin{equation}
  M M^\dag = U_R \Lambda_R (U_R)^\dagger \ ,
\end{equation}
where $\Lambda_R$ is a diagonal matrix whose diagonal elements are
the eigenvalues of $M M^\dagger$.
We also obtain $U_L$ for the left side in the same way.
We can evaluate the truncation error $\epsilon_R$ and $\epsilon_L$ for $U_R$ and $U_L$:
\begin{equation}
  \epsilon_{R(L)} = \sum_{i>D} (\Lambda_{R(L)})_{ii} \ .
\end{equation}
We adopt the one with the smaller truncation error between $U_R$ and $U_L$ as $U_x$.
$U_y$ and $U_z$ for the $y$ and $z$ directions are obtained in the same way:
$M = A^{(0)} \otimes A^{(1)} \otimes B$ for the $y$ direction,
and $M = A^{(1)} \otimes A^{(2)} \otimes B$ for the $z$ direction.
Finally, we obtain the initial tensor $T$ by contracting $A^{(0)}$, $A^{(1)}$, $A^{(2)}$, $B$, $B$, $B$, $U_x$, $U_y$, and $U_z$ as in Fig.~\ref{fig: HOSVD}.


\bibliographystyle{ptephy_arxiv}
\bibliography{ref}

\end{document}